\documentclass[prd,superscriptaddress]{revtex4}
\usepackage{graphicx}
\usepackage{epsfig}
\DeclareGraphicsExtensions{.pdf}
\usepackage{amssymb}
\usepackage{appendix}
\usepackage{amsmath}  

\makeatletter

\newcommand{\Rmnum}[1]{\expandafter\@slowromancap\romannumeral #1@}    
\makeatother

\begin{document}

\title{New targets for relic antineutrino capture}

\author{Jeong-Yeon Lee}
\email{yeon.lee@ssu.ac.kr} 
\affiliation{Origin of Matter and Evolution of Galaxies (OMEG) Institute, Soongsil University,  Seoul 06978, Republic of Korea}

\author{Yeongduk Kim}
\email{ydkim@ibs.re.kr}
\affiliation{Center for Underground Physics, Institute for Basic Science, Daejeon 34126, Republic of Korea}

\author{Satoshi Chiba}
\email{chiba.satoshi@nr.titech.ac.jp}
\affiliation{Laboratory for Advanced Nuclear Energy, Tokyo Institute of Technology, Meguro, Tokyo 152-8550, Japan}

\begin{abstract}

$^{163}$Ho has been considered as a suitable candidate for the capture of relic antineutrinos.  
However,  the detection of the relic antineutrino using $^{163}$Ho is extremely challenging with current techniques. 
Therefore, we have searched for new targets for relic antineutrino detections through the resonant capture on nuclides undergoing electron capture. 
We have investigated nuclear and atomic properties of all nuclides. And we finally propose  $^{131}$Ba,  $^{159}$Dy, $^{175}$Hf, $^{195}$Au, and $^{243}$Cm as new candidates for the relic antineutrino detection, and call for high precise experiments of $Q_{\rm EC}$-values and intensities of EC decays for these new candidates. 
  
\end{abstract}

\maketitle

\section{Introduction}

As well as the 2.7 K microwave background radiation, the universe is filled with a sea of relic neutrinos that decoupled from the rest of the universe within the first 10 s after the Big Bang.  
The average number density of the neutrinos  in the Universe is  $\sim$55$~{\rm cm}^{-3}$  for neutrinos (or antineutrinos) of each flavor.  
Still, nobody has detected these relic neutrinos yet. 
The overdensity ($=\frac{\rho(x)-\bar{\rho}}{\bar{\rho}}$)  in the neutrino distribution around the Earth is expected to be less than $\sim$10$^{4}$, which is the matter overdensity in our galaxy~\cite{Rin04}, and may be as small as 3~\cite{Sal17}. 
These relic neutrinos may have played a crucial role in the nucleosynthesis, structure formation, and the evolution of the universe as a whole.   
A non-zero neutrino mass can dramatically change  the properties of the relic neutrino sea and its role in the evolution of the universe. 
The current cosmological average neutrino mass is $\sim$0.15 eV~\cite{ Sal17} or tightly constrained to $<$0.12 eV~\cite{Pla18}. 

In principle relic neutrinos could be detected via capture on unstable nuclei and the finite mass has an effect in the kinematics of the capture process and leads to the possibility, at least in principle, to unambiguously detect the very low energy cosmological neutrino backgrounds~\cite{Irv83,Coc07}. 
Low energy electron capture (EC) and  beta decay are excellent tools for testing non-zero neutrino masses.   
Among all beta unstable nuclides,  $^{3}$H($Q_{\beta^{-}}$=18.5906 32 keV)~\cite{Kat05,Pur15,Mye15,Asn15},  $^{187}$Re($Q_{\beta^{-}}$=2.4670 20 keV )~\cite{Bas09},  
and $^{163}$Ho($Q_{\rm EC}$=2.833(30$_{\rm stat}$)(15$_{\rm sys}$) keV)~\cite{Eli15,Alp15,Gas17} are attractive for non-zero neutrino mass experiments due to their low $Q$-values. 

EC decaying $^{163}$Ho with low $Q_{\rm EC}$-value has been proposed as a candidate for measuring the electron neutrino mass since a long time~\cite{Ruj82}, and interest in  $^{163}$Ho has been renewed by Lusignoli and Vignati~\cite{Lus11} and Li and Xing~\cite{Li11} with a possible relic antineutrino capture experiment.  
According to their calculation~\cite{Lus11}, taking the  $Q_{\rm EC}$=2.8 keV, 1274 kg $^{163}$Ho  is necessary for observing relic antineutrino capture,  and this amount is many orders of magnitude larger than 5.6$\times$10$^{-7}$ kg corresponding to 10 MBq planned for direct neutrino mass determination experiments. 
In 2014, Vergados and Novikov have published a paper suggesting that $^{157}$Tb can be a better candidate than $^{163}$Ho for a relic antineutrino capture experiment with prospects of detection of relic antineutrinos by resonant absorption in electron capturing nuclei~\cite{Ver14}.  
However, as will be explained in detail in section 2, required $^{157}$Tb mass for the experiment enormously exceeds what can be practically available.  

The small number of relic antineutrino capture signal can  be observed on the right-hand side of the endpoint of the EC spectrum. Detector featuring an energy resolution 
at the eV level or below will be necessary to avoid or at least reduce the overlap of the tiny resonance corresponding to the captured relic antineutrino with the tail of the EC
spectrum. In addition, an extremely low background needs to be achieved so that the number of events related to the capture of relic antineutrino would be statistically relevant. 
The KATRIN neutrino mass experiment has the possibility to observe relic neutrino capture on tritium if the relic neutrino overdensity is greater than 
 2$\times$10$^{9}$~\cite{Kab10},  which is several orders of magnitude larger than the expected overdensity.   
Therefore, the necessary target mass for relic antineutrino detection using  $^{163}$Ho or $^{157}$Tb  is many orders of magnitude larger than what can be realistically considered for an experiment with the required energy resolution and low background level. 

The difficulty in measuring the relic antineutrinos using the $^{163}$Ho and $^{157}$Tb  calls for new candidate targets for the experiments. 
In this work, we have considered all nuclei with nuclear data available and searched for new targets for detection of the relic antineutrino, using the resonant absorption in electron capturing nuclei. 
Section 2 shows a brief description of electron capture and relic antineutrino capture, and the expected signature in the measured spectra due to the capture of relic antineutrinos along with a summary of previous works. 
Section 3 describes the search for new targets for relic antineutrino capture and present the nuclides which feature most suitable properties to be used in the experiments.
Section 4 presents conclusions. 

\section{Electron capture and relic antineutrino capture}

In an EC unstable atom, a bound electron can be captured by the nucleus, leaving a hole in an electron state of the daughter atom and releasing a neutrino.  
The de-excitation of the excited atom can occur via photon and/or Auger electron emission, which can be measured.
The EC decay rate can be obtained by~\cite{Lus11,Bam77} 
\begin{equation}
\lambda_{\rm EC} =\frac {G_{\beta}^{2}}{4{\pi}^{2}} \sum\limits_{i}n_{i} C_{i} \beta_{i}^{2}B_{i}(Q-E_{i}) \sqrt{(Q-E_{i})^2 - m_{\nu}^{2}}   
\label{ECDecayRate}
\end{equation}
with $G_{\beta}$=$G_{F}$cos${\rm \theta_{C}}$, $n_{i}$ the fraction of occupancy of the $i$-th atomic shell, $C_{i}$ the nuclear shape factor, 
and $B_{i}$ the correction for electron exchange and overlap. 
$\beta_{i}$   is the amplitude of the electron radial wave function  and 
represents the overlap of the electron wave function with the nucleus.  
Since the EC decay rate is proportional to the electron density at the nucleus~\cite{Bam77}, captures from the $\rm P_{3/2}$ electron states do not contribute to the reaction rate and the EC can occur, in first approximation, only from $\rm S$ and $\rm P_{1/2}$ electron states. 
As shown with the high resolution measurements of the $^{163}$Ho spectra by the ECHo collaboration~\cite{Gas14}, this approximation is correct at $\sim$5 \% level. 
More detailed description of the electron capture in $^{163}$Ho is given in the papers by Brass {\it et al.}~\cite{Bra18,Bra20}.   

A very important information to select suitable nuclei for cosmic neutrino background detection is a precise knowledge of the $Q$-value. 
High precision Penning trap mass spectroscopy (PT-MS)~\cite{Eli15,Bla10} can accurately measure the mass difference between the parent and the daughter atoms, and this technique was used for direct measurement of the mass difference of $^{163}$Ho and $^{163}$Dy for solving the $Q$-value puzzle which was of utmost importance for planning neutrino mass experiments using $^{163}$Ho~\cite{Eli15}. Another key aspect towards an experiment for the detection 
of cosmic neutrino background is a detector technology which ensures high energy resolution and that is applicable to relatively large masses.
If the source is enclosed in the detector, both electron and photons emitted in the atomic de-excitation process can be detected~\cite{Gas13,Gat97}. 
In case of the decay to a nuclear excited state, the gamma emitted during the transition to the ground stated should be then detected with a different detector to allow for a coincidence measurement. 

Atomic levels have finite  natural widths and the spectrum of calorimetric energy $E_{c}$ can be given by~\cite{Cuo04}
\begin{equation}
\frac {d\lambda_{\rm EC}}{dE_{c}} =\frac {G_{\beta}^{2}}{4{\pi}^{2}} (Q-E_{c}) \sqrt{ (Q-E_{c})^{2} - m_{\nu}^{2}}~   \sum\limits_{i} n_{i} C_{i} \beta_{i}^{2}B_{i}
\frac{\Gamma_i}{2\pi} \frac{1}{{(E_{c}-E_{i})^2} + \Gamma_{i}^{2}/4}.    
\label{spectrum}
\end{equation}  
This neutrino spectrum is the complementary spectrum with respect to the de-excitation spectrum of the daughter atom.   
Recently, the EC de-excitation spectrum of $^{163}$Ho has been measured with metallic magnetic calorimeters with high resolution of $\Delta E_{\rm FWHM}$=8.3 eV~\cite{Ran17}.  

Since the crossed reaction with an incoming antineutrino  has no threshold energy~\cite{Coc07}, a nucleus can absorb a very low energy  antineutrino  and an electron from a bound $i$-th atomic shell, which is called antineutrino capture. 
The antineutrino capture rate can be obtained by~\cite{Lus11} 
\begin{equation}
\lambda_{\bar{\nu}} =n_{\bar{\nu}} \frac {G_{\beta}^{2}}{2} \sum n_{i} \beta_{i}^{2} B_{i} \rho_{i}(E_{\bar{\nu}}),   
\label{ReactionRate}
\end{equation}
where $E_{\bar{\nu}}$ ($\simeq m_{\nu}$  for C$\nu$B) is the energy of incoming antineutrinos,  $\rho_{i}(E_{\bar{\nu}})$  is the density of final states, 
and $n_{\bar{\nu}}(\sim$55~${\rm cm}^{-3})$ is the number density of the antineutrinos which, without over density is considered.    

The ratio between the antineutrino capture and the EC decay rates  can be obtained by~\cite{Lus11}   
\begin{equation}
\frac{\lambda_{\bar{\nu}}}{\lambda_{\rm EC}} \simeq 2\pi^{2}  n_{\bar{\nu}} 
\frac {\sum_{i} n_{i} \beta_{i}^{2} B_{i} \rho_{i}(E_{\bar{\nu}})}       
{\sum_{i} n_{i} \beta_{i}^{2} B_{i} (Q_{\rm EC}-E_{i})^2} .
\label{Ratio} 
\end{equation} 
The total number of signal events can be calculated using~\cite{Lus11} 
\begin{equation}
S =   \frac{\lambda_{\bar{\nu}}}{\lambda_{\rm EC}} \frac{{\rm log} 2}{T_{1/2}} N_{A}n_{\rm mol}t, 
\label{NoEvent}
\end{equation} 
where $N_A$ is Avogadro's number, $n_{\rm mol}$ is the number of moles, $t$ is the exposure time, and  $T_{1/2}$ is the half-life of the parent nucleus.   
The relic antineutrino capture signal should be seen on the right-hand side of the spectral endpoint of the EC decay, and their interval should be detectable, provided the target is big enough and the energy resolution is good enough. 

Assuming  $Q_{\rm EC}$=2.5 keV for $^{163}$Ho, Lusignoli and Vignati obtained the ratio of the rates of the two processes,   ${\lambda_{\bar{\nu}}}/{\lambda_{\rm EC}}$=5.8$\times$10$^{-23}$~\cite{Lus11}, which is   higher than the analogous result for tritium $\beta$-decay, $\lambda_{\nu}/{\lambda_{\beta}}$=6.6$\times$10$^{-24}$~\cite{Coc07}.  
This ratio corresponds to 307 kg  of $^{163}$Ho target to get 10 events of signal per year for a relic antineutrino capture experiment. 
However, considering a recent experimental result  $Q_{\rm EC}$=2.833(30$_{\rm stat}$)(15$_{\rm sys}$) keV for $^{163}$Ho~\cite{Eli15}  
and assuming the most optimistic   $Q_{\rm EC}^{\rm opt}$=2.788 keV, 
we have obtained   ${\lambda_{\bar{\nu}}}/{\lambda_{\rm EC}}$=1.45$\times$10$^{-23}$ which indicates that we need 1.23$\times$10$^{3}$ kg of $^{163}$Ho target for 10 events per year.  
This amount of target can be compared with the current ECHo and HOLMES experiments using the $^{163}$Ho, though these experiments are for neutrino mass measurement, not relic neutrino detection. The final activity for the ECHo experiment is $\sim$1 MBq (56 $\mu $g). HOLMES experiment will be similar to ECHOs.
This implies the fact that at the present knowledge experiments for relic antineutrino detections using $^{163}$Ho target are beyond feasibility.  

In~\cite{Ver14}, Vergados and Novikov claimed that $^{157}$Tb can be a better target than $^{163}$Ho  with prospects of detection of relic antineutrinos by       
introducing the resonance mechanism of the relic antineutrino absorption on EC decaying nuclides. 
The resonant absorption of the relic antineutrino becomes possible when the atomic mass difference is equal to the binding energy of one of the captured electrons, specifically when the excitation energy of the daughter state is equal to the relic antineutrino total energy, and if the resonant conditions for the antineutrino absorption are met, a considerable enhancement of the absorption rates can be obtained, which makes the relic antineutrino detections more promising.   
However,  the current  $Q_{\rm EC}$=60.044$\pm$0.297 keV~\cite{Wan21} for $^{157}$Tb does not allow the resonant absorption of an antineutrino,  and 
assuming the most optimistic  $Q_{\rm EC}^{\rm opt}$=59.747 keV  for  $^{157}$Tb decaying to the 54.53 keV excited state of $^{157}$Gd, 
we have obtained the ratio 8.70$\times 10^{-26}$ which corresponds to 2.79$\times 10^{6}$ kg of $^{157}$Tb source for 10 events per year. 
The amount of target source required when using $^{163}$Ho or $^{157}$Tb is many orders of magnitude greater than what current experiments can deal with, and the difficulty  calls for new targets for the experiment, and finding the new targets is the purpose of this research.

\section{Search for new target for relic antineutrino capture and results}

 \begin{figure}
\begin{center}
\includegraphics[scale=0.45]{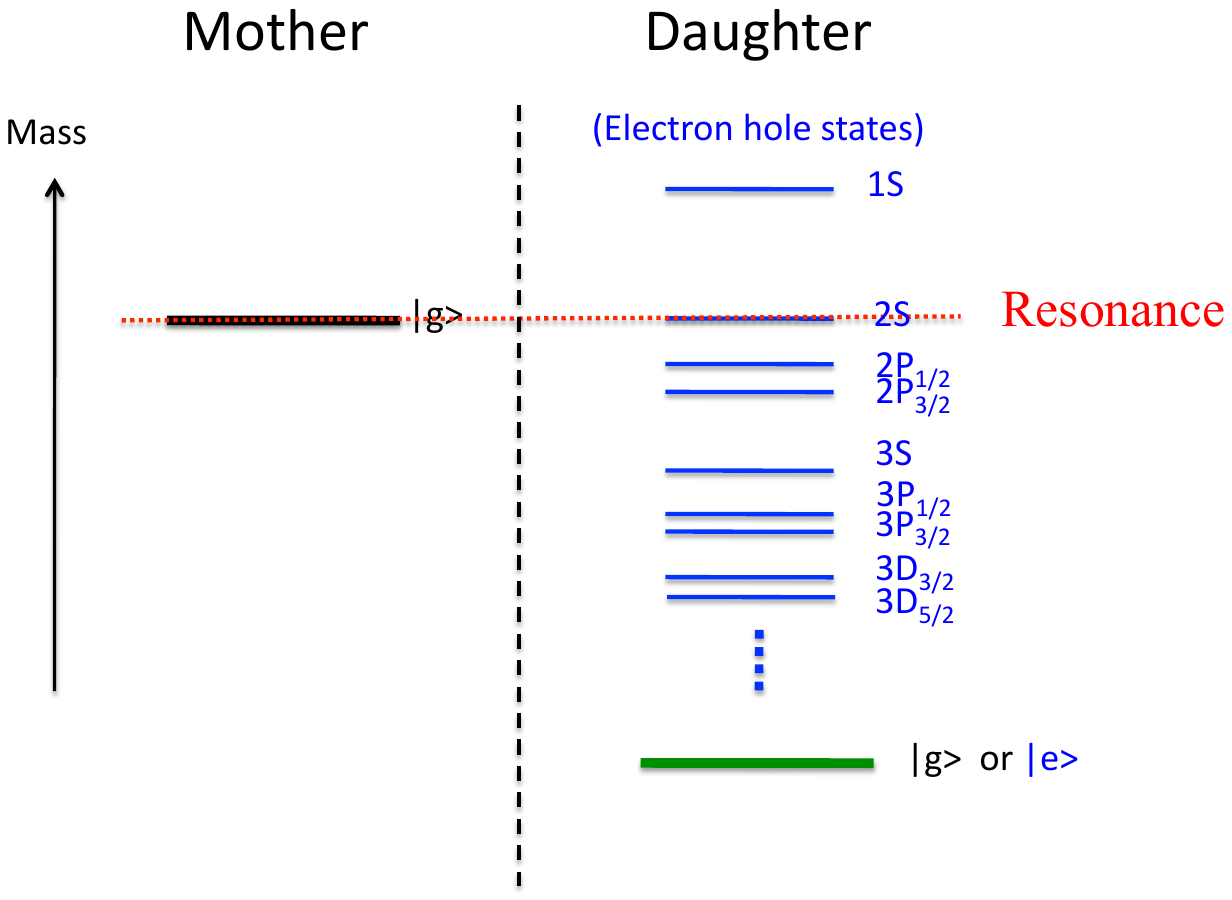}
 \caption{Concept of the resonant EC decay. When $Q_{\rm EC}$-value matches the excitation energy of the daughter, there can be resonant absorption of electron and antineutrino.}
\label{levelscheme}        
\end{center}
 \end{figure}
 
A large $Q_{\rm EC}$-value and  a large distance to atomic excitation resonances imply a strong suppression of the capture rate, leading to a worse signature of the relic antineutrino in the experiment. 
Therefore, the nuclides of  smaller $Q_{\rm EC}$-values are  eagerly wanted, in order to do a more promising experiment for the direct detection of the relic antineutrino.   
To search for new targets for relic antineutrino detections, we use the resonance mechanism of capture/absorption on nuclides undergoing electron capture~\cite{Ver14}.  
A schematic concept is shown in Fig.~\ref{levelscheme}. 
When $Q_{\rm EC}$-value matches the excitation energy of the daughter, 
whether decayed to the nuclear ground state or to an excited state, there can be the resonant absorption of an antineutrino on the EC decaying nuclides.  
When the resonant absorption  occurs, the capture rate is greatly increased, thereby enabling the difficult experiment with a smaller amount.

A very important information to select suitable nuclides for the relic antineutrino detection is a precise knowledge of the $Q_{\rm EC}$-value. 
However, compared to the errors in the electron binding energies and the nuclear excited state energies, the errors in $Q_{\rm EC}$ that come from the mass excesses of the parent and daughter nuclei are very large. 
We newly define $Q'_{\rm EC}$ for an EC decay to a daughter excited state      
\begin{eqnarray}
Q'_{\rm EC} & = &\Delta m_{(Z)} - \Delta m_{(Z-1)}^{* \dagger}  \\
             & = &\Delta m_{(Z)} - (E_{(Z-1).nucl} + \Delta m_{(Z-1)}^{\dagger}) \\ 
             & = &\Delta m_{(Z)} - (E_{(Z-1).nucl} + \Delta m_{(Z-1)}+ B_{(Z-1).e}- B_{(Z-1).e.val})   
\label{QValue} 
\end{eqnarray} 
with $\Delta m$ denoting the mass excess, '$\dagger$' atomic excitation,  '*' nuclear excitation, 
$E_{(Z-1).nucl}$ the nuclear excitation energy  of the daughter nucleus with proton number $'Z-1'$, 
$B_{(Z-1).e}$ the binding energy of the electron being captured, 
and $B_{(Z-1).e.val}$ the binding energy of the valence electron of the daughter atom. 
We define  $dQ'_{\rm EC}$ and  $\Delta Q'_{\rm EC}$ as 
\begin{equation}
dQ'_{\rm EC}  = \sqrt{\Delta m_{(Z)}^2 +\Delta m_{(Z-1)}^{2}}, ~~~ \Delta Q'_{\rm EC}  = |Q'_{\rm EC}| -  dQ'_{\rm EC}.  
\label{DeltaQ}  
\end{equation}      
$dQ'_{\rm EC}$ represents the uncertainty in $Q'_{\rm EC}$ and comes from the uncertainties in mass excesses. 
$\Delta Q'_{\rm EC}$ represents the degree of uncertainty in $Q'_{\rm EC}$-value.  
In case $\Delta Q'_{\rm EC} \le 0 $, $Q'_{\rm EC}$-value  for a resonant absorption lies in the error bar in $Q_{\rm EC}$-value and therefore 
we can expect a resonant absorption of an antineutrino as well as an electron.

\begin{table}
\centering
\caption{Results for candidate nuclei decaying to the ground states of their daughters: required amounts of targets and the corresponding activities to get 10 events per year for relic antineutrino detection for the new candidate $^{243}$Cm and the previously suggested  $^{163}$Ho.   $^{243}$Cm and $^{163}$Ho decay to their daughter ground states with  intensities of EC,  $ I_{\epsilon}$. 
}
\begin{tabular}{ccc|cccc|c}   \hline \hline
Parent  & $J^{\pi}$  & $ T_{1/2}$      & Daughter   &  $J^{\pi}$  &  $\!\!T_{1/2}$ &  $ I_{\epsilon}$  & $Q_{\rm EC}$  \\ 
    &      &     (y)    &    &     &   (y) & (\%)&   (keV)    \\  \hline
  $^{163}$Ho &  7/2$^{-}$   & 4570  & $^{163}$Dy   &  5/2$^{-}$    &  stable & 100     & 2.833(30stat)(15sys)     \\  \hline 
  $^{243}$Cm & 5/2$^{+}$  &  29.1  &  $^{243}$Am    &  5/2$^{-}$    &  7370 & 0.29       &    7.473 $\pm$ 1.713   \\
\hline \hline
\label{candidate-gs}              
\end{tabular}
\begin{tabular}{c|ccc|cc} \hline \hline
 Parent      &   Res.state & $Q_{\rm EC}^{\rm opt}$  &  ${\lambda_{\bar{\nu}}}/{\lambda_{\rm EC}} $ & Amount  &  Activity \\
       &      & (keV)                                                &                           &  (kg)  & (MCi)    \\  \hline
  $^{163}$Ho &   -  & 2.788 & 1.45$\times 10^{-23}$    & 1.23$\times 10^{+3}$ & 5.89$\times 10^{-1}$    \\  \hline 
  $^{243}$Cm & 3S(M1) &6.133 & 3.42$\times 10^{-20}$  & 1.71$\times 10^{+0}$   & 8.64$\times 10^{-2}$  \\
\hline \hline
\end{tabular}
\end{table}               

To search for suitable targets for relic antineutrino detection, 
we consider all nuclides with nuclear data~\cite{Eli15,Wan21,ensdf} and atomic data~\cite{CRC14,Ban86}  available and 
set up the following criteria: 
(a) A nucleus should be EC unstable. 
(b) $\Delta J=$1 to be of allowed beta decays. 
(c) $T_{1/2}$ of the parent nucleus is longer than 10 days for realistic experiments. 
(d) $Q_{\rm EC}, Q'_{\rm EC}, dQ'_{\rm EC}   \le 10 $ keV. 
(e) $\Delta Q'_{\rm EC}   \le 0 $ keV.  

We calculate $Q_{\rm EC}$, $Q'_{\rm EC}$, $dQ'_{\rm EC}$, and $\Delta Q'_{\rm EC}$ and impose the above criteria to sort out possible candidates from all nuclei available. 
We scrutinize  the sorted out candidates and find the most optimistic $Q_{\rm EC}$-values, $Q_{\rm EC}^{\rm opt}$,  which allow the resonant absorptions of antineutrinos, 
within the $Q_{\rm EC}$ or $Q'_{\rm EC}$. 
Assuming these $Q_{\rm EC}^{\rm opt}$-values, 
we calculate the ratios ${\lambda_{\bar{\nu}}}/{\lambda_{\rm EC}}$ using Eq.~\ref{Ratio},  required amounts of targets using Eq.~\ref{NoEvent}, and their activities to get 10 events per year. 

We first focus on EC isotopes with low $Q_{\rm EC}$, decaying to the 'ground' states of the daughters. 
Beside the well known candidate $^{163}$Ho ($Q_{\rm EC}$=2.833(30$_{stat}$)(15$_{sys}$) keV, $T_{1/2}$=4570 y)~\cite{Eli15}, $^{243}$Cm ($Q_{\rm EC}$=7.473$\pm$1.713 keV, $T_{1/2}$=29.1 y)~\cite{Wan21}  has appeared as a new prospective target.  
The $^{243}$Cm decays only in 0.29 \% via EC and predominantly via alpha emission with a branching ratio of 99.71\%  and therefore the background due to the highly emitting alphas should be well treated. 
$Q_{\rm EC}^{\rm opt}$ of $^{243}$Cm is 6.133 keV, which corresponds to a resonant capture from 3S(M1) electron state of binding energy 6.133 keV, and we have obtained the highest ratio ${\lambda_{\bar{\nu}}}/{\lambda_{\rm EC}}$=3.42$\times 10^{-20}$ for  $^{243}$Cm.  
In this case, we need only 1.71 kg of $^{243}$Cm source, which is 3 orders of magnitude less than 1.23$\times$10$^{3}$ kg of the $^{163}$Ho mentioned above. The calculated results  are shown in Table~\ref{candidate-gs} with experimental data~\cite{Eli15,Wan21,ensdf}. 

\begin{figure}
\begin{center}
  \includegraphics[scale=0.60]{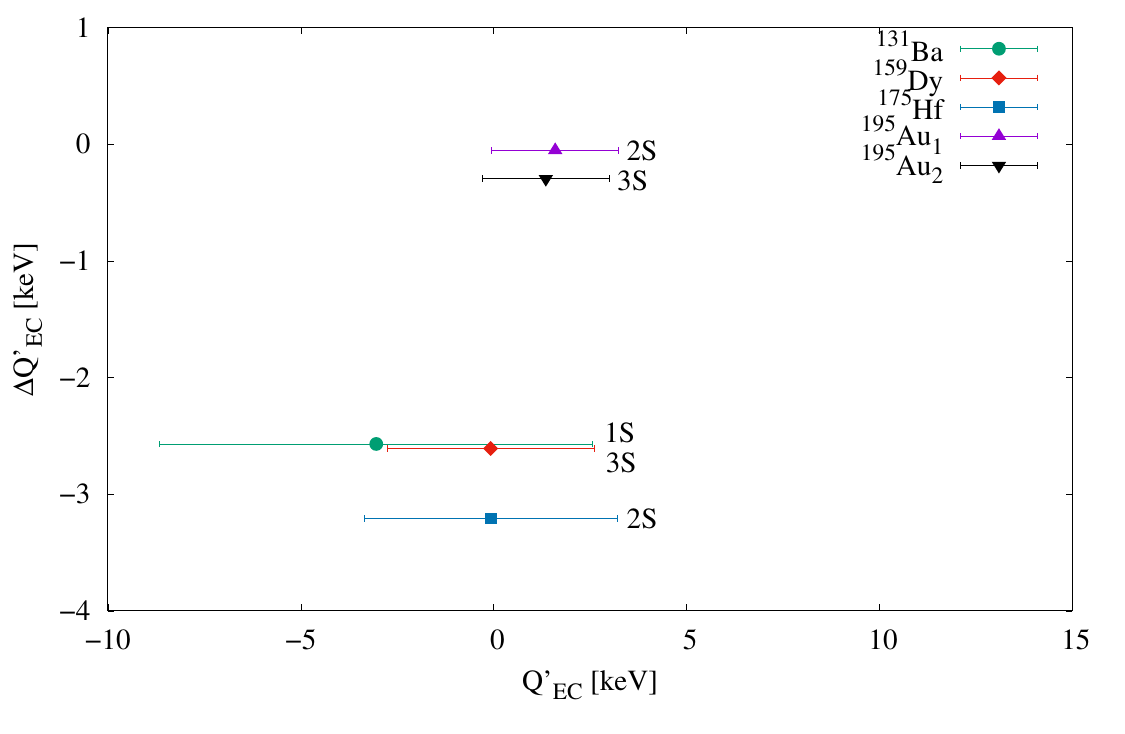}
 \vspace{-0.6cm}
 \caption{New target nuclides decaying via EC to excited nuclear states. $^{195}$Au$_{1}$ and $^{195}$Au$_{2}$ represent for decay to the 211.4 keV and 222.2 keV excited states of $^{195}$Pt, respectively. Electrons in the states given in the figure can be resonantly absorbed by the nuclei.}
\label{4relic-error}        
\end{center} 
 \end{figure}

For EC decays to ’excited’ states, we have found $^{131}$Ba,  $^{159}$Dy, $^{175}$Hf, and $^{195}$Au as new prospective candidate nuclei. 
The candidate nuclei are shown in Fig.~\ref{4relic-error} in terms of $Q'_{\rm EC}$ and $\Delta Q'_{\rm EC}$. 
In the figure, we define $^{195}$Au$_{1}$ and $^{195}$Au$_{2}$ to represent for decay to the 211.4 keV and 222.2 keV excited states of $^{195}$Pt, respectively. 
Negative $\Delta Q'_{\rm EC}$ means that $Q'_{\rm EC}$-value, allowing  a resonant absorption, lies in the error bar in $Q_{\rm EC}$-value, and the electrons in the states shown in the figure can be resonantly captured.
For $^{131}$Ba,  $^{175}$Hf,  and $^{195}$Au$_{2}$, the information of the EC intensity $ I_{\epsilon}$ of the daughter excited nuclear state is missing and therefore the calculations of the required amounts cannot be done. However, $^{131}$Ba,  $^{175}$Hf,  and $^{195}$Au$_{2}$ have high ratios of the rates of the two processes and furthermore $Q_{\rm EC}^{\rm opt}$  for $^{131}$Ba corresponds to a resonant capture from 1S(K) state with large amplitude of the electron wave function  $\beta_{i}$, and therefore, according to their measured EC intensities, the detector can require similar or even smaller amounts of the nuclides compared with the other suggested targets, and the $^{131}$Ba,  $^{175}$Hf,  and $^{195}$Au$_{2}$ nuclides can be prospective target candidates. 
We call for measurements of EC intensity $ I_{\epsilon}$ for the prospective candidates for precise calculations of the rates of the relic antineutrino captures on the nuclides against the corresponding EC decay rates. 
Presently among the studied nuclides decay via electron capture to excited nuclear states, $^{195}$Au$_{1}$  and $^{159}$Dy are the ones which look most promising. 
Assuming the most optimistic $Q_{\rm EC}^{\rm opt}$-values which allow the resonant absorptions, 
we need only  1.44 kg of $^{159}$Dy and 0.82 kg of $^{195}$Au$_{1}$ sources for 10 events per year, while we need 
1.23$\times$10$^{3}$ kg of  $^{163}$Ho and 2.79$\times$10$^{6}$ kg of $^{157}$Tb.  

\begin{figure}
 \begin{center}
\includegraphics[scale=0.60]{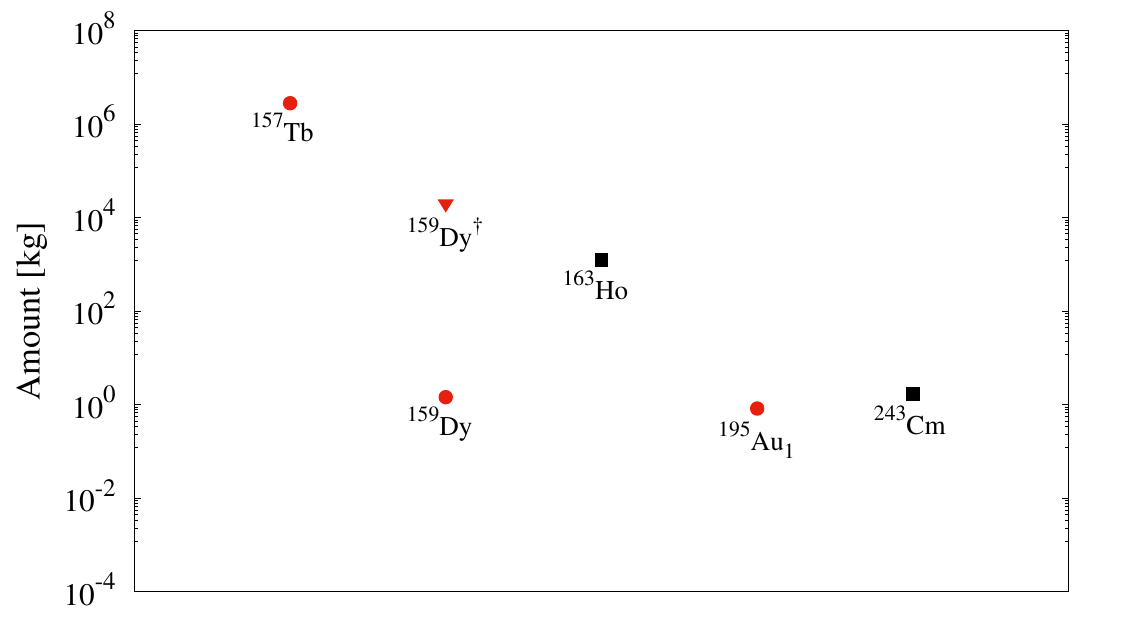}
 \includegraphics[scale=0.60]{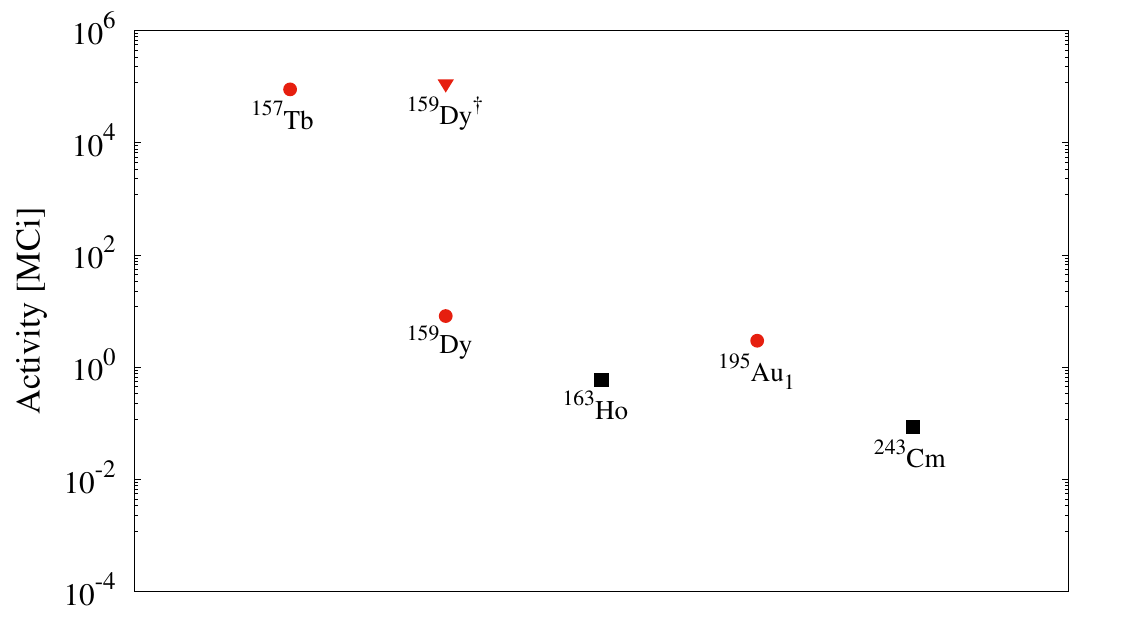}
 \caption{Required amounts of the sources (upper) and the corresponding activities (lower) for 10 events per year for new candidates  $^{159}$Dy,  $^{195}$Au$_{1}$, and $^{243}$Cm, and the previously suggested candidates $^{163}$Ho and $^{157}$Tb. Square and round dots denote nuclei decaying  to ground and excited states of the daughter nuclei, respectively.
 Also shown are the results for $^{159}$Dy$^{\dagger}$ using the recently measured $Q_{\rm EC}$-value~\cite{Ge21}.}  
\label{candidates-new}        
\end{center}
\end{figure}

The calculated results for the new prospective nuclei decaying to excited states  are shown in Table~\ref{candidate-ex} with experimental data~\cite{Wan21,ensdf}. 
We emphasize that the event rate related to the ratio ${\lambda_{\bar{\nu}}}/{\lambda_{\rm EC}}$ is very sensitive to the $Q_{\rm EC}$-value and  highly precise measurements of $Q_{\rm EC}$ are required for the new targets we propose.   
Figure~\ref{candidates-new} displays required amounts of the target sources (upper) and the corresponding activities (lower) to get 10 events per year for our new prospective targets  $^{159}$Dy,  $^{195}$Au$_{1}$, and $^{243}$Cm and the previously suggested candidates $^{163}$Ho and $^{157}$Tb. 
Square and round dots in the figure denote for  decays  to the ground and excited states of daughter nuclei, respectively.   

\begin{table} 
\centering
\caption{Results for candidate nuclei decaying to excited states of their daughters:  required amounts of targets and  the corresponding activities to get 10 events per year  for relic antineutrino capture experiments . Also shown are the results for the $^{157}$Tb proposed by Vergados and Novikov~\cite{Ver14} and the $^{159}$Dy$^{\dagger}$ using the most recent experimental $Q_{\rm EC}$-value~\cite{Ge21}. } 
\begin{tabular}{lcc|cccc|c} \hline \hline
~~ Parent            &\shortstack{$J^{\pi}$ }         & $ T_{1/2}$                  & Daughter  $\!\!\!$    & $E_{\rm ex}$ & $J^{\pi}$ & $ I_{\epsilon}$  & $Q_{\rm EC}$   \\
                                                &                      &     (s)                             &                                &  (keV)    &                   & (\%)       &   (keV)  \\  \hline
 ~~$^{157}$Tb                       &  3/2$^{+}$    & 2.239$\times 10^{9}$   &  $^{157}$Gd          &  54.53  & 5/2$^{-}$   & 0.11       & 60.044$\pm$0.297    \\  \hline 
 ~~$^{131}$Ba                       &  1/2$^{+}$    & 9.936$\times 10^{5}$   & $^{131}$Cs            &   1342. & 3/2$^{+}$  & -             & 1375.958$\pm$5.287  \\ 
 ~~$^{175}$Hf                       &  5/2$^{-}$     & 6.048$\times 10^{6}$   & $^{175}$Lu            &  672.9  & 7/2$^-$      &  -            & 683.712$\pm$0.195   \\ 
 ~~$^{195}$Au$_{1}$           &  3/2$^{+}$    & 1.608$\times 10^{7}$  &  $^{195}$Pt$_{1}$ & 211.4   & 3/2$^{-}$   & 0.0195   & 226.8$\pm$1.0   \\ 
 ~~$^{195}$Au$_{2}$           &  3/2$^{+}$    & 1.608$\times 10^{7}$  &  $^{195}$Pt$_{2}$ & 222.2   & 1/2$^{-}$   & -             &  226.8$\pm$1.0  \\ \hline
 ~~$^{159}$Dy                      &  3/2$^{-}$     & 1.248$\times 10^{7}$   & $^{159}$Tb           &  363.5   & 5/2$^{-}$  & 0.00019 & 365.400$\pm$1.200   \\ 
 ~~$^{159}$Dy$^{\dagger}$ &  3/2$^{-}$    & 1.248$\times 10^{7}$   & $^{159}$Tb           &  363.5   & 5/2$^{-}$  & 0.00019 & ~~~364.73$\pm$0.19$^{\dagger}$\cite{Ge21} \\
\hline \hline  
\label{candidate-ex}   
\end{tabular}
\begin{tabular}{l|rr|crc|cc} \hline \hline
~~ Parent          & $Q'_{\rm EC}$   & $\Delta Q'_{\rm EC}$ & Res.state & $Q_{\rm EC}^{\rm opt}$ & ${\lambda_{\bar{\nu}}}/{\lambda_{\rm EC}}$ & Amount & Activity \\
                        &     (keV)   &  (keV)                       &     &  (keV)                             &                           & $\!\!$(kg)  & (MCi) \\ \hline 
~~ $^{157}$Tb & 1.714      & $-$0.600                  & - & 59.747                        & 8.70$\times 10^{-26}$   & 2.79$\times 10^{+6}$  &     8.95$\times 10^{+4}$ \\ \hline 
~~ $^{131}$Ba &    $-$3.043  &   $-$2.569            &   1S(K)   &   1377.945                     & 2.76$\times 10^{-21}$      &  -   ~~~~                               &  -    ~~           \\
~~ $^{175}$Hf &    $-$0.070   & $-$3.208              &  2S(L1)   & 683.770                    & 4.12$\times 10^{-20}$      & -  ~~~~                                 & -  ~~ \\
~~ $^{195}$Au$_{1}$ &   1.596        &   $-$0.053             &  2S(L1)    & 225.286                     &  1.48$\times 10^{-20}$       & 8.24$\times 10^{-1}$ & 2.97$\times 10^{+0}$    \\
~~ $^{195}$Au$_{2}$ &   1.353        &   $-$0.295             & 3S(M1)   & 225.526                     & 1.69$\times 10^{-19}$       & -    ~~~ ~                              &  - ~~\\ \hline
~~ $^{159}$Dy &    $-$0.078   & $-$2.608              & 3S(M1)    & 365.512                     &  5.52$\times 10^{-19}$     & 1.44$\times 10^{+0}$  & 8.16$\times 10^{+0}$    \\
~~ $^{159}$Dy$^{\dagger}$ &   - ~~  & -  ~~        &  -  &364.54     ~&  3.95$\times 10^{-23}$     & 2.01$\times 10^{+4}$  & 1.14$\times 10^{+5}$ \\    
\hline \hline
\end{tabular}
\end{table}

 \begin{figure}
\begin{center}
\includegraphics[scale=0.53]{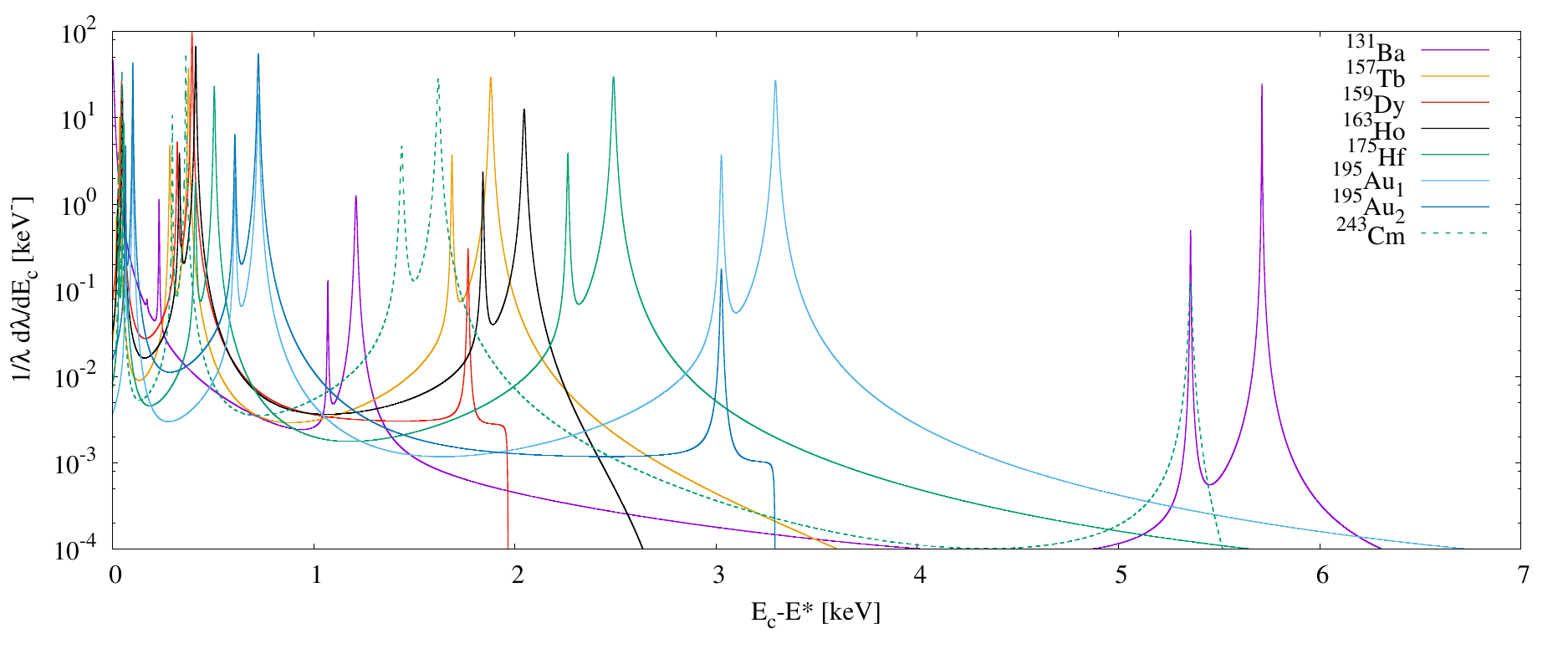}
\caption{Expected de-excitation energy spectra of EC decays of candidates for  relic antineutrino capture experiments. } 
\label{spect-all-m02}        
\end{center}
 \end{figure}

\begin{figure}
\begin{center}
\includegraphics[scale=0.60]{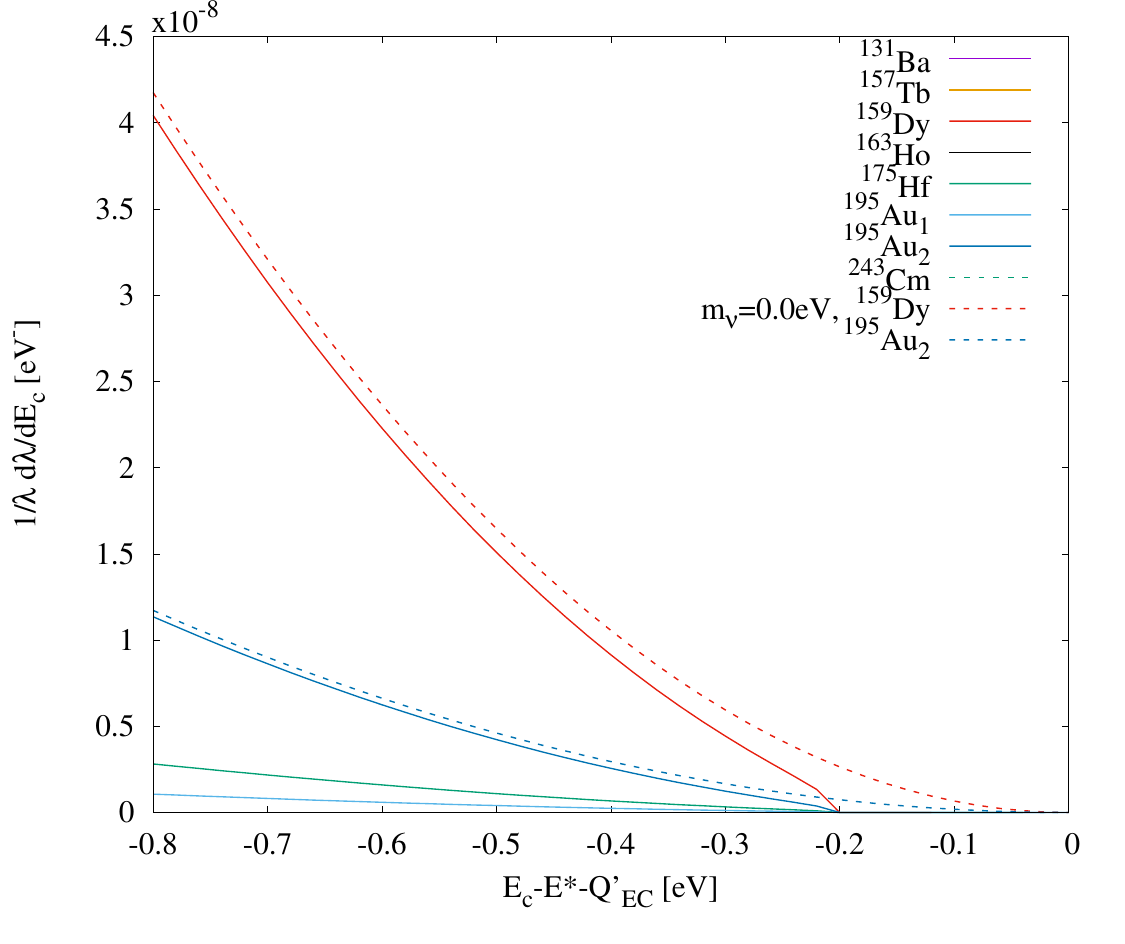}
\caption{Expected de-excitation energy spectra of EC decays of the candidate targets in their tails. The signature of the relic antineutrino capture is  located on the right-hand side of the spectral endpoint.} 
\label{spect-all-tail}        
\end{center}
 \end{figure}

Recent measurement of $Q_{\rm EC}$ on $^{159}$Dy  using a double Penning trap mass spectrometer gives 364.73(19) keV with high precision~\cite{Ge21}. 
This new  result does not allow the resonant absorption of an antineutrino and,  in the case of  $Q_{\rm EC}^{\rm opt}$=365.43 keV,  the  ratio  becomes  
3.95$\times 10^{-23}$ which corresponds to 2.01$\times 10^{4}$ kg of  $^{159}$Dy and activity 1.14$\times 10^{5}$ MCi for 10 events per year. 
The calculated results using this new measurement on $^{159}$Dy  are also shown in Table~\ref{candidate-ex} and  
Fig.~\ref{candidates-new} with a  '$\dagger$' symbol and compared with the ones using the AME2020 data~\cite{Wan21}. 
In case we consider this new experimental result 364.73(19) keV, we may need to delist the $^{159}$Dy from the candidate targets. 

The de-excitation energy spectra of the EC decays are shown in Fig.~\ref{spect-all-m02} as a function of '$E_{C}-E^*$' for the candidates, considering  the  neutrino mass $m_{\nu}$=0.2 eV. 
Also the tail parts of the de-excitation energy spectra for the candidates are shown in  Fig.~\ref{spect-all-tail} and compared with the ones of $m_{\nu}$=0 eV cases for the targets $^{159}$Dy and  $^{195}$Au$_{2}$ which have larger values in the tail part.    
The end point spectrum is very sensitive to the electron neutrino mass, and the neutrino mass can be more easily determined  using the nuclei which give higher count rates in the tail part.  
The signature of the relic antineutrino capture is located on the right-hand side of the spectral endpoint of the EC decay, and the antineutrino capture signal can be well separated, depending on the amount of target and the energy resolution but also on the fact that the major resonances have a distance which is many times larger than the energy resolution. 
The distance between the peak of the antineutrino capture signature and the EC decay background becomes larger for a larger neutrino mass, and accordingly the required energy resolution becomes less stringent.

 $^{159}$Dy and $^{175}$Hf  can be produced by thermal neutron capture by enriched $^{158}$Dy and $^{174}$Hf,  
which have  large cross sections of 43(6) and 549(7) barns, respectively~\cite{Mug18}. 
Irradiating thermal neutron flux higher than ${\rm 10^{13} neutrons/cm^{2}/sec}$  on highly enriched $^{174}$Hf target, 
we can obtain kilograms of $^{175}$Hf per year.  
However, at this moment, considering the state of the art of the detector in terms of energy resolution and multiplexing readout, it is not trivial to predict how radioactive material should be enclosed in detector which can host just enough atoms to get a few Bq. 
And the practical detection of the relic antineutrino seems to be very challenging even with our new targets $^{159}$Dy,  $^{195}$Au, and $^{243}$Cm which require orders of magnitude less amounts of targets  than the previously suggested nuclei $^{163}$Ho and $^{157}$Tb.

\section{Conclusions}

The universe is filled with relic neutrinos and still nobody has yet detected these relic neutrinos.  
EC decaying $^{163}$Ho and $^{157}$Tb nuclides with low $Q_{\rm EC}$-values have been suggested as candidates 
for measuring the relic antineutrinos. 
However, the capture rate of the antineutrino on these nuclei is so small that the desired target mass must be formidably large. 
We have  investigated the nuclear and atomic properties of all nuclides to search for new targets for relic antineutrino capture experiments using the resonant mechanism of the relic antineutrino absorption in EC decaying nuclei.  

We finally propose the nuclei $^{131}$Ba,  $^{159}$Dy, $^{175}$Hf, $^{195}$Au, and $^{243}$Cm as  new prospective targets and call for highly precise measurements of the $Q_{\rm EC}$-values for  these new candidate targets,  as well as the intensities of EC decays to the daughter excited nuclear states for $^{131}$Ba and $^{175}$Hf.   
However, the recent experimental result on $^{159}$Dy, $Q_{\rm EC}$=364.73(19) keV~\cite{Ge21}, does not cover the $Q_{\rm EC}$=365.512 keV for the resonant absorption of an antineutrino, and considering this new experimental data, we may delist the $^{159}$Dy from the candidate targets. 
Considering current technology in detectors, the practical detection of the relic antineutrino seems to be very challenging even using our new prospective targets,  which require orders of magnitude less amounts of targets  than the previously suggested nuclei $^{163}$Ho and $^{157}$Tb.

\end{document}